\documentclass[prc,preprintnumbers,onecolumn]{revtex4}
\usepackage{graphicx}
\usepackage{amssymb}
\usepackage{amsmath}

\begin{document}

\title{Derivation of capture cross section from  quasielastic excitation function}
\author{V.V.Sargsyan$^{1,2}$, G.G.Adamian$^{1}$, N.V.Antonenko$^{1}$, and P.R.S.Gomes$^3$}
\affiliation{$^{1}$Joint Institute for Nuclear Research, 141980 Dubna, Russia\\
$^{2}$International Center for Advanced Studies, Yerevan State University, 0025 Yerevan, Armenia\\
$^{3}$Instituto de Fisica, Universidade Federal Fluminense, Av. Litor\^anea, s/n, Niter\'oi, R.J. 24210-340, Brazil}

\date{\today}

\begin{abstract}
The relationship between  the quasielastic excitation function
and the capture cross section is  derived.
The quasielastic data
is shown to be a useful tool to extract the  capture cross sections and the angular momenta
of the captured systems for the reactions $^{16}$O+$^{144,154}$Sm,$^{208}$Pb, $^{20}$Ne+$^{208}$Pb, and
$^{32}$S+$^{90,96}$Zr
at  near and above the Coulomb barrier energies.
\end{abstract}

\pacs{25.70.Jj, 24.10.-i, 24.60.-k \\ Key words: capture cross section,
quasielastic excitation function, cold fusion reactions}

 \maketitle

\section{Introduction}
The partial capture cross section is  one of the important ingredients to calculate and predict
the
production cross sections of  exotic and superheavy nuclei in the
cold, hot, and sub-barrier astrophysical fusion reactions.
Therefore, more experimental and theoretical
studies of the capture  process are required.
There is a relationship between the capture and
the quasielastic scattering processes because of the conservation
of the reaction flux~\cite{PRSGomes1,PRSGomes3}.
Any loss from
the quasielastic channel  directly
contributes to the capture and vise versa.
The quasielastic measurements are usually not as complex as the direct capture (fusion)
measurements. Thus, the quasielastic data are  suited for the extraction of
the capture probabilities and of
the capture cross sections.

The paper is organized in the following way. In Sec.~II we derive the formulas for
the extraction of the capture cross section and of the angular momentum of the captured system
by employing the
experimental quasielastic excitation function. In Sec.~III, using these formulas,
we extract the capture cross sections and the angular momenta
of the captured systems and compare  with  those of direct measurements.
Using the available experimental
quasielasic data, we  predict the capture cross sections
for the cold fusion reactions. In Sec.~IV the paper is summarized.

\section{Relationship between capture and quasielastic scattering}
The expression
\begin{equation}
P_{qe }(E_{\rm c.m.},J) + P_{cap}(E_{\rm c.m.},J)=1
\end{equation}
connecting the quasielastic (reflection) $P_{qe}$  and
 the capture (transmission) $P_{cap}$ probabilities
follows from the conservation of the reaction flux~\cite{PRSGomes1,PRSGomes3}.
Thus, one can extract the capture probability $P_{cap}(E_{\rm c.m.},J=0)$ at $J=0$ from
the experimental quasielastic probability $P_{qe}(E_{\rm c.m.},J=0)$:
\begin{equation}
P_{cap}(E_{\rm c.m.},J=0)=1-P_{qe}(E_{\rm c.m.},J=0)=1-d\sigma_{qe}(E_{\rm c.m.})/d\sigma_{Ru}(E_{\rm c.m.}).
\end{equation}
Here,
the quasielastic probability~\cite{PRSGomes1,Timmers,Timmers2,Zhang}
\begin{eqnarray}
P_{qe}(E_{\rm c.m.},J=0)=d\sigma_{qe}/d\sigma_{Ru}
\end{eqnarray}
for
angular momentum $J=0$ is given by the ratio of
the quasielastic differential cross section  and
Rutherford differential cross section at 180 degrees.
Further, one can approximate the $J$ dependence of the capture probability $P_{cap}(E_{\rm c.m.},J)$
at a given energy $E_{\rm c.m.}$ by  shifting the energy~\cite{Bala}:
\begin{eqnarray}
P_{cap}(E_{\rm c.m.},J)\approx P_{cap}(E_{\rm c.m.}-\frac{\hbar^2\Lambda }{2\mu R_b^2},J=0)
=1-P_{qe}(E_{\rm c.m.}-\frac{\hbar^2\Lambda }{2\mu R_b^2},J=0),
\end{eqnarray}
where $\Lambda=J(J+1)$, $R_b=R_b(J=0)$ is the position of the Coulomb barrier at $J=0$.
Then, we extract the capture cross section $\sigma_{cap}(E_{\rm c.m.})$ from the
experimental quasielastic probabilities $P_{qe}$:
\begin{eqnarray}
\sigma_{cap}(E_{\rm c.m.})=\sum_{J=0}^{J_{cr}}\sigma_{\rm cap}(E_{\rm
c.m.},J)=
\pi\lambdabar^2
\sum_{J=0}^{J_{cr}}(2J+1)[1-P_{qe}(E_{\rm c.m.}-\frac{\hbar^2\Lambda }{2\mu R_b^2},J=0)],
\label{1a_eq}
\end{eqnarray}
where $\lambdabar^2=\hbar^2/(2\mu E_{\rm c.m.})$ is the reduced de Broglie wavelength,
$\mu=m_0A_1A_2/(A_1+A_2)$ is the reduced mass ($m_0$ is the nucleon mass),
and at  given bombarding energy $E_{\rm c.m.}$
the summation is over the possible values of angular momentum $J$ from $J=0$ to
the critical angular momentum $J=J_{cr}$.
For values $J$ greater than $J_{cr}$, the potential pocket in the
nucleus-nucleus interaction potential vanishes and the capture is not occur.
To calculate the critical angular momentum $J_{cr}$ and
 the position $R_{b}$
of the Coulomb barrier,
we use the nucleus-nucleus interaction potential $V(R,J)$ of  Ref.~\cite{Pot}.
For the nuclear part of the nucleus-nucleus potential, the double-folding formalism with
the Skyrme-type density-dependent effective nucleon-nucleon interaction is employed~\cite{Pot}.

If one sets $R_b(J)\approx R_b$ in Eq.~(5) for approximating the $J$-wave penetrability
by the $s$-wave penetrability at a shifted energy, one obtains only the leading term in
the series expansion in $\Lambda$. The next term in this
expansion can be easily calculated in the same way as in Ref.~\cite{Bala}
[$R_b(J)\approx R_b-\frac{\hbar^2\Lambda}{\mu\alpha R_b^3}$,
$V_b(J)\approx V_b+\frac{\hbar^2\Lambda}{2\mu R_b^2}+\frac{\hbar^4\Lambda^2}{2\mu^2\alpha R_b^6}$,
$\alpha=-\partial^2 V(R,J=0)/\partial R^2|_{R=R_b}=\mu\omega_b^2$, $\omega_b=\omega_b(J=0)$ is the
curvature of the $s$-wave potential barrier with the height $V_b=V_b(J=0)=V(R=R_b,J=0)$]:
\begin{eqnarray}
P_{cap}(E_{\rm c.m.},J)\approx P_{cap}(E_{\rm c.m.}-
\frac{\hbar^2\Lambda}{2\mu R_b^2}-\frac{\hbar^4\Lambda^2}{2\mu^2\alpha R_b^6},J=0).
\end{eqnarray}
With this improved expression for the $P_{cap}$, we obtain
\begin{eqnarray}
\sigma_{cap}(E_{\rm c.m.})= \pi\lambdabar^2
\sum_{J=0}^{J_{cr}}(2J+1)[1-P_{qe}(E_{\rm c.m.}-\frac{\hbar^2\Lambda }{2\mu R_b^2},J=0)][1-\frac{2\hbar^2\Lambda }{\mu^2\omega_b^2 R_b^4}].
\label{1a2_eq}
\end{eqnarray}
Converting the sum over $J$ into an integral and changing variables to
$E=E_{\rm c.m.}-\frac{\hbar^2\Lambda }{2\mu R_b^2}$ in Eq.~(7),
we obtain the following simple expression:
\begin{eqnarray}
\sigma_{cap}(E_{\rm c.m.})
=\frac{\pi R_b^2}{E_{\rm c.m.}}
\int_{E_{\rm c.m.}-\frac{\hbar^2\Lambda_{cr}}{2\mu R_b^2}}^{E_{\rm c.m.}}dE[1-d\sigma_{qe}(E)/d\sigma_{Ru}(E)][1-\frac{4(E_{\rm c.m.}-E)}{\mu\omega_b^2  R_b^2}],
\label{3cn_eq}
\end{eqnarray}
which relates the capture cross section with quasielastic excitation function.
Note that $\Lambda$ is
not a small parameter, there is a natural cutoff $\Lambda_{cr}=J_{cr}(J_{cr}+1)$ in this parameter. Because
of this cutoff, the second term $\frac{\hbar^2\Lambda}{2\mu R_b^2}$
in Eq.~(6) is always larger than the third one $\frac{\hbar^4\Lambda^2}{2\mu^2\alpha R_b^6}$~\cite{Bala}.
By using the experimental quasielastic probabilities $P_{qe}(E_{\rm c.m.},J=0)$ and Eq.~(8)
one can obtain the capture cross sections.

For the systems with $Z_1\times Z_2 < 2000$, the critical angular momentum $J_{cr}$ is large enough
and Eqs.~(7) and (8) can be approximated with a good accuracy as:
\begin{eqnarray}
\sigma_{cap}(E_{\rm c.m.})\approx \pi\lambdabar^2
\sum_{J=0}^{\infty }(2J+1)[1-P_{qe}(E_{\rm c.m.}-\frac{\hbar^2\Lambda }{2\mu R_b^2},J=0)][1-\frac{2\hbar^2\Lambda }{\mu^2\omega_b^2 R_b^4}]
\label{1a2x_eq}
\end{eqnarray}
and
\begin{eqnarray}
\sigma_{cap}(E_{\rm c.m.})
\approx \frac{\pi R_b^2}{E_{\rm c.m.}}
\int_{0}^{E_{\rm c.m.}}dE[1-d\sigma_{qe}(E)/d\sigma_{Ru}(E)][1-\frac{4(E_{\rm c.m.}-E)}{\mu\omega_b^2  R_b^2}].
\label{3cx_eq}
\end{eqnarray}

Following the procedure of Ref.~\cite{Bala} and using the extracted $\sigma_{cap}$ and the experimental
$P_{qe}$, one can find the average angular momentum
\begin{eqnarray}
<J>
=\frac{\pi R_b^2}{E_{\rm c.m.}\sigma_{cap}(E_{\rm c.m.})}
\int_{E_{\rm c.m.}-\frac{\hbar^2\Lambda_{cr}}{2\mu R_b^2}}^{E_{\rm c.m.}}
&&dE[1-d\sigma_{qe}(E)/d\sigma_{Ru}(E)]
[1-\frac{5(E_{\rm c.m.}-E)}{\mu\omega_b^2 R_b^2}]\nonumber\\
&&\times [(\frac{2\mu R_b^2}{\hbar^2}(E_{\rm c.m.}-E)+\frac{1}{4})^{1/2}-\frac{1}{2}]
\label{3cxJ_eq}
\end{eqnarray}
and the second moment of the angular momentum
\begin{eqnarray}
<J(J+1)>
=\frac{2\pi\mu R_b^4}{\hbar^2E_{\rm c.m.}\sigma_{cap}(E_{\rm c.m.})}
\int_{E_{\rm c.m.}-\frac{\hbar^2\Lambda_{cr}}{2\mu R_b^2}}^{E_{\rm c.m.}}
&&dE[1-d\sigma_{qe}(E)/d\sigma_{Ru}(E)]
[1-\frac{6(E_{\rm c.m.}-E)}{\mu\omega_b^2 R_b^2}]\nonumber\\
&&\times [E_{\rm c.m.}-E]
\label{3cxJ2_eq}
\end{eqnarray}
of the captured system.

\section{Results of calculations}
For the verification of our method of the extraction of $\sigma_{cap}$,
firstly we compare the extracted
capture cross sections with experimental one.
In Figs.~1 and 2 one can see a good agreement between the extracted and directly
measured capture cross sections for the reactions $^{16}$O + $^{120}$Sn, $^{18}$O + $^{124}$Sn,
$^{16}$O + $^{208}$Pb, and $^{16}$O + $^{144}$Sm at energies above the Coulomb barrier.
The results on the sub-barrier energy region are discussed later on.
To extract the capture cross section, we use both Eq.~(8) (solid lines) and Eq.~(10) (dotted lines).
The used values of critical angular momentum are $J_{cr}$=54, 56, 57, and 62
for the reactions $^{16}$O + $^{120}$Sn, $^{18}$O + $^{124}$Sn,
$^{16}$O + $^{144}$Sm, and $^{16}$O + $^{208}$Pb, respectively.
The difference between the
results of Eqs.~(8) and (10)  is less than 5$\%$ at the highest
energies. At low energies,  Eqs.~(8) and (10) lead to the same values
of  $\sigma_{cap}$. The factor $1-\frac{4(E_{\rm c.m.}-E)}{\mu\omega_b^2  R_b^2}$ in  Eqs.~(8) and (10)
very weakly influences the results of the  calculations for the systems and energies considered.
Hence, one can say that for the relatively light systems the proposed method of extracting the capture cross section
is model independent (particular, independent on the potential used).

\begin{figure}
\includegraphics[scale=1]{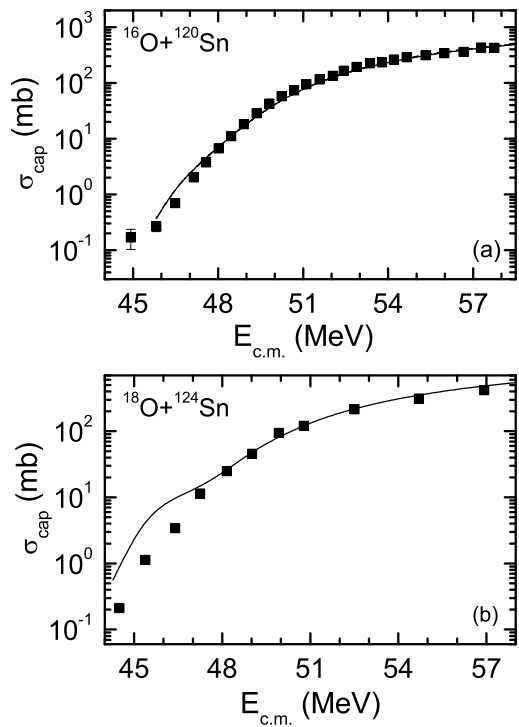}
\caption{
The extracted capture cross sections for the  reactions $^{16}$O + $^{120}$Sn (a) and  $^{18}$O + $^{124}$Sn (b)
by employing
Eq.~(8) (solid line) and Eq.~(10) (dotted line). These lines are almost coincide.
The used experimental quasielastic data are from Ref.~\protect\cite{Sinha}.
The experimental capture (fusion) data (symbols) are from Refs.~\protect\cite{Sinha,JACOBS}.
}
\label{1_fig}
\end{figure}
\begin{figure}
\includegraphics[scale=1]{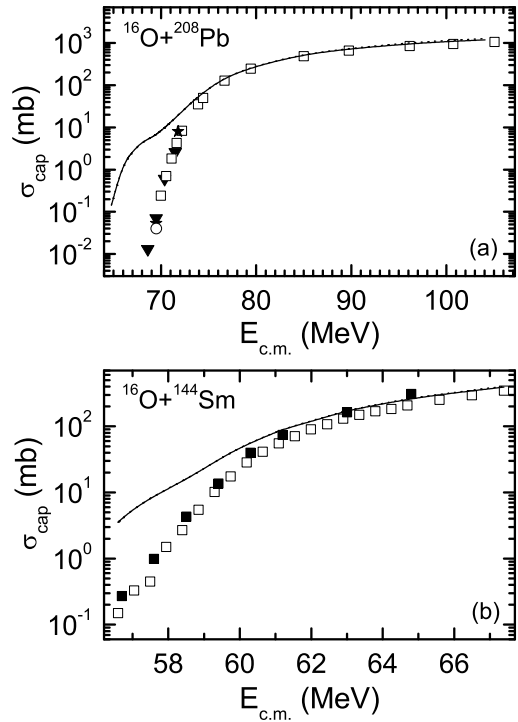}
\caption{The same as in Fig.~1, but  for the reactions
$^{16}$O + $^{208}$Pb(a),$^{144}$Sm(b).
The used  experimental quasielastic data are from Refs.~\protect\cite{Timmers2,Timmers}.
For the  $^{16}$O + $^{208}$Pb reaction, the experimental capture (fusion) data are from
Refs.~\protect\cite{Pbcap} (open squares),~\protect\cite{Pbcap1} (open circles),~\protect\cite{Pbcap2} (closed stars),
and~\protect\cite{Pbcap3} (closed triangles).
For the $^{16}$O + $^{144}$Sm  reaction, the experimental capture (fusion) data  are
from Refs.~\protect\cite{SmCap1} (closed squares) and~\protect\cite{SmCap2} (open squares).
}
\label{2_fig}
\end{figure}
One can see that the used formulas are suitable not only for  almost spherical nuclei
(Figs.~1~and~2), but also for the reactions
with strongly deformed target- or projectile-nucleus (Figs.~3~and~4).
The deformation effect is effectively contained  in the experimental $P_{qe}$.
$J_{cr}=58$,
 68, 74, and 76 for the reactions  $^{16}$O+$^{154}$Sm,
$^{32}$S+$^{90}$Zr,  $^{32}$S+$^{96}$Zr, and $^{20}$Ne+$^{208}$Pb, respectively.
The results obtained by employing the formula (10) are almost the same and not presented in Figs.~3~and~4.
\begin{figure}
\includegraphics[scale=1]{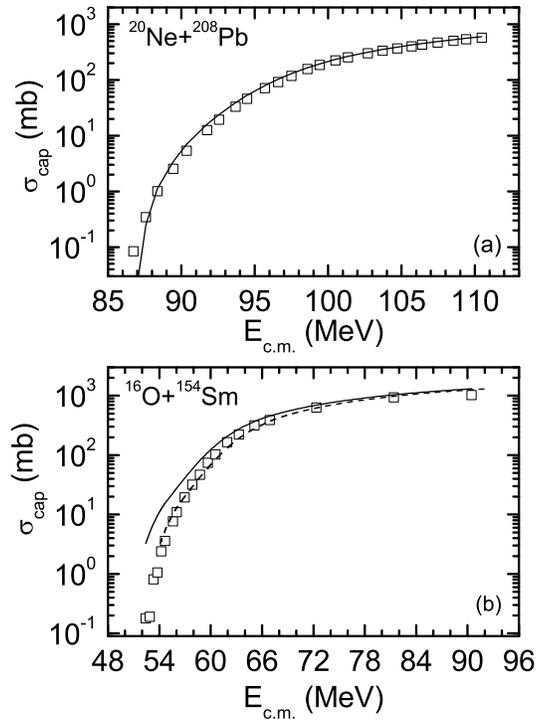}
\caption{
The same as in Fig.~1, but for the  reactions $^{20}$Ne + $^{208}$Pb and $^{16}$O + $^{154}$Sm.
The used  experimental quasielastic data are from Refs.~\protect\cite{Piasecki,Timmers}.
The experimental capture (fusion) data (symbols) are from Refs.~\protect\cite{SmCap2,Piasecki}.
For the  $^{16}$O + $^{154}$Sm reaction,
the dashed line is obtained from the shift of the solid line  by  1.7 MeV higher energies.
}
\label{3_fig}
\end{figure}

\begin{figure}
\includegraphics[scale=1]{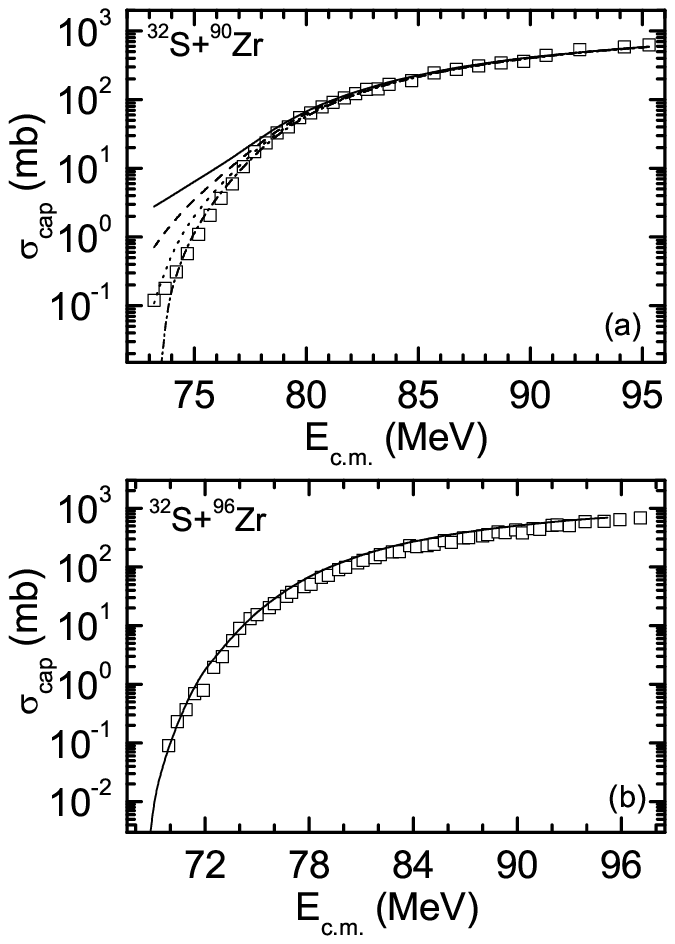}
\caption{
The same as in Fig.~1, but for the  reactions $^{32}$S + $^{90}$Zr (a) and $^{32}$S + $^{96}$Zr (b).
For the  $^{32}$S+$^{90}$Zr   reaction, we show the extracted capture cross sections,
increasing  the experimental $P_{qe}$ by 1\% (dashed line), 2\% (dotted line), and 3\% (dash-dotted line).
The  used experimental quasielastic data are from Ref.~\protect\cite{Zhang3}.
The experimental capture (fusion) data (symbols) are from Ref.~\protect\cite{ZhangS32Zn9096}.
For the $^{32}$S + $^{96}$Zr  reaction, the energy scale for the extracted capture cross sections
is adjusted to that of direct measurements.
}
\label{4_fig}
\end{figure}
For the reactions $^{16}$O+$^{154}$Sm and  $^{32}$S+$^{96}$Zr,
the extracted capture cross sections are shifted in energy by 1.7 and 1.9 MeV, respectively, with respect to the measured capture data.
This could be
the result of different energy calibrations in the experiments on the capture measurement and on the
quasielastic scattering. Because of the lack of systematics in these energy shifts, their
origin remains unclear and we adjust  the
Coulomb barriers in the extracted capture cross sections to the values following the experiments.

Note that the extracted
and experimental capture cross sections deviate from each other in the reactions
$^{16}$O+$^{208}$Pb, $^{16}$O+$^{144}$Sm, and  $^{32}$S+$^{90}$Zr at energies below the Coulomb barrier.
Probably this deviation is a reason for the large discrepancies 
in the diffuseness parameter extracted from the analyses 
of the quasielastic scattering and fusion (capture) at deep sub-barrier energies.
One of the possible reasons for the overestimation of the capture cross section from the quasielastic data
at sub-barrier energies is
the underestimation of the total reaction differential  cross section taken
as the Rutherford differential cross section.
Indeed, for the  $^{32}$S+$^{90}$Zr   reaction,
the increase of $P_{qe}$ within 2--3\% is in order to obtain
the agreement between the extracted and measured
capture cross sections at the sub-barrier energies~[Fig.~4(a)].

One can use  Eq.~(8) and available experimental
quasielasic data~\cite{Ikezoe}  to predict the capture cross sections
for the reactions $^{48}$Ti,$^{54}$Cr,$^{56}$Fe,$^{64}$Ni,$^{70}$Zn + $^{208}$Pb,
using
$J_{cr}=78$, 74, 58, 51, 31, respectively.
The extracted capture cross sections $\sigma_{cap}(E_{\rm c.m.})$ as a function of $E_{\rm c.m.}$
are presented in Fig. 5 (a).
\begin{figure}
\includegraphics[scale=1]{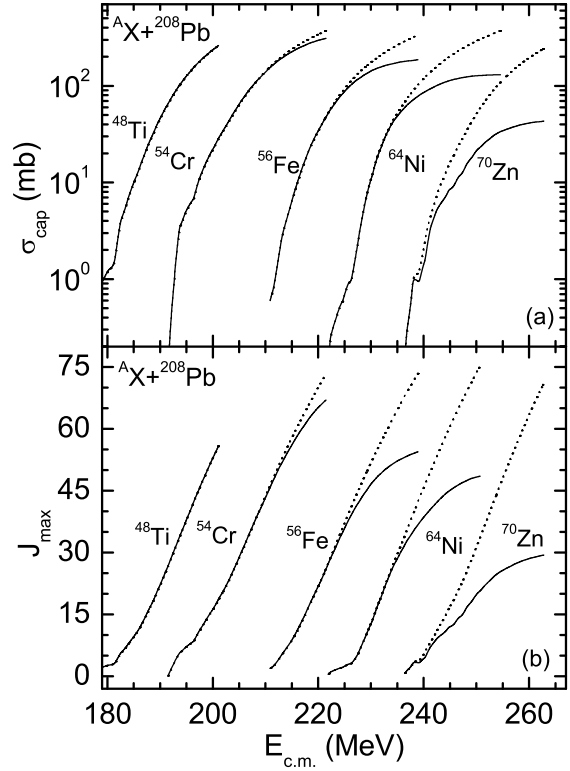}
\caption{(a) The extracted capture cross sections employing  Eq.~(8) (solid line) and
Eq.~(10) (dotted line) for the reactions
$^{48}$Ti,$^{54}$Cr,$^{56}$Fe,$^{64}$Ni,$^{70}$Zn + $^{208}$Pb.
The used experimental quasielastic data are from Ref.~\protect\cite{Ikezoe}.
(b) The extracted values of the maximal angular momenta vs.
energy for the above mentioned reactions.
The solid and dotted lines  show the results of calculations of  $J_{max}$ by using  the
extracted capture cross sections  calculated with  Eqs.~(8) and (10), respectively.
}
\label{5_fig}
\end{figure}
The formulas~(8)~and~(10) give almost the same capture cross sections for
reactions $^{48}$Ti,$^{54}$Cr + $^{208}$Pb at energies under consideration. Thus, for these systems,
the values of $J_{cr}$ are relatively large and the account of $J_{cr}$ does not affect the results.
However, for  heavier systems with smaller $J_{cr}$ (the smaller potential pockets in the
nucleus-nucleus interaction potentials), the deviation between the results obtained with Eqs.~(8)~and~(10)
increases  strongly with the factor $Z_1\times Z_2$.
The $\sigma_{cap}$, calculated with the finite value of critical
angular momentum, decreases with increasing  Coulomb repulsion in the system.
One can try  to check experimentally these predictions of $\sigma_{cap}(E_{\rm c.m.})$
by the direct measurement of the capture cross sections.
Note that the values of the extracted capture cross sections for the  $^{48}$Ti + $^{208}$Pb system are
close to those found in the experiments $^{50}$Ti + $^{208}$Pb \cite{Naik,Clerc}.
However, for the $^{64}$Ni + $^{208}$Pb system,
there are strong deviations in the energy between the extracted
and experimental~\cite{Bock} capture cross sections.

By using the extracted $\sigma_{cap}(E_{\rm c.m.})$ and
the sharp-cutoff approximation, one can determine the maximal angular momentum $J_{max}$
in the captured system as a function of the bombarding energies:
\begin{eqnarray}
J_{max}=[2\mu E_{\rm c.m.}\sigma_{cap}(E_{\rm c.m.})/(\pi\hbar^2)]^{1/2}-1.
\label{lcrit}
\end{eqnarray}
The extracted $J_{max}$ for the cold fusion reactions are shown in Fig.~5(b).
For the system $^{70}$Zn + $^{208}$Pb, the small depth of the potential pocket in the
nucleus-nucleus interaction potential  leads to the decrease of
$J_{max}$  by the factor about of 2.4 at highest energy considered (about of 17 MeV above the Coulomb barrier).

In the reactions
with weakly bound nuclei one can extract the capture cross section by employing
the conservation of the reaction flux~\cite{PRSGomes1,Nash,Be9Pb,Li6Pb}
\begin{eqnarray}
P_{cap}(E_{\rm c.m.},J=0)=1-[P_{qe}(E_{\rm c.m.},J=0)+P_{BU}(E_{\rm c.m.},J=0)]
\label{Pcapbreakup}
\end{eqnarray}
and
the measured probabilities of the quasielastic scattering ($P_{qe}(E_{\rm c.m.},J=0)=d\sigma_{qe}/d\sigma_{Ru}$) and
of the breakup ($P_{BU}(E_{\rm c.m.},J=0)=d\sigma_{BU}/d\sigma_{Ru}$) which are defined as the differential
cross sections ratios between quasielastic scattering, breakup reaction and the
Rutherford scattering at backward angle.
As seen in Fig.~6,
the extracted capture cross sections $\sigma_{cap}(E_{\rm c.m.})$ (solid line)
for the $^{6}$Li+$^{208}$Pb reaction are
rather close to those found in the direct measurements~\cite{Li6Pbcap}
at energies above the Coulomb barrier.
\begin{figure}
\includegraphics[scale=1]{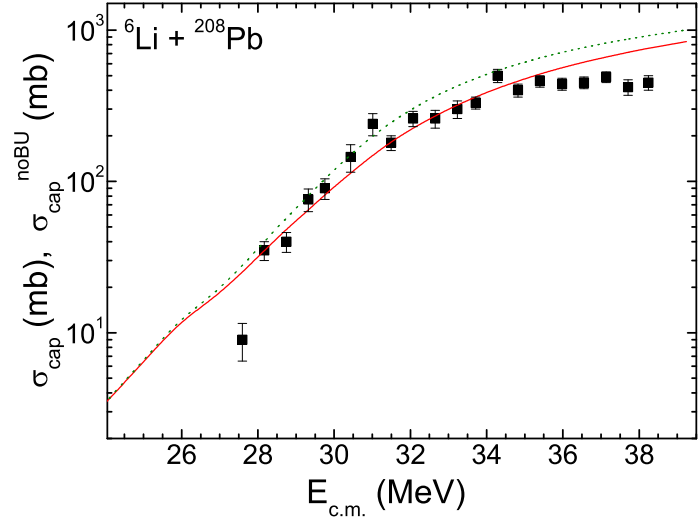}
\caption{
(Colour online)
The extracted capture cross sections $\sigma_{cap}(E_{\rm c.m.})$ (solid line)  and
$\sigma^{noBU}_{cap}(E_{\rm c.m.})$  (dotted line)
for the $^{6}$Li+$^{208}$Pb reaction.
The used experimental quasielastic and quasielastic plus breakup data are from Ref.~\protect\cite{Li6Pb}.
The experimental capture cross sections (solid squares)  are from Refs.~\protect\cite{Li6Pbcap}.
The energy scale for the extracted capture cross sections
is adjusted to that of direct measurements.
}
\label{6_fig}
\end{figure}
It looks that at energies near and below the Coulomb barrier
 the extracted $\sigma_{cap}(E_{\rm c.m.})$ deviates from the direct measurements.
It is similarly possible to calculate the capture excitation function
\begin{eqnarray}
\sigma^{noBU}_{cap}(E_{\rm c.m.})
=\frac{\pi R_b^2}{E_{\rm c.m.}}
\int_{E_{\rm c.m.}-\frac{\hbar^2\Lambda_{cr}}{2\mu R_b^2}}^{E_{\rm c.m.}}dEP^{nBU}_{cap}(E,J=0)[1-\frac{4(E_{\rm c.m.}-E)}{\mu\omega_b^2  R_b^2}]
\label{noBU}
\end{eqnarray}
in the absence of the breakup
process (Fig.~6, dotted line) by using the following formula for the capture
probability in this case~\cite{Nash}:
\begin{eqnarray}
P^{nBU}_{cap}(E_{\rm c.m.},J=0)=1-\frac{P_{qe}(E_{\rm c.m.},J=0)}{1-P_{BU}(E_{\rm c.m.},J=0)}.
\label{P_noBU}
\end{eqnarray}
By employing the  measured excitation functions $P_{qe}$ and
$P_{BU}$ at backward angle~\cite{Li6Pb}, Eqs.~(8),~(15), and the
formula
\begin{eqnarray}
<P_{BU}>(E_{\rm c.m.})=1-\frac{\sigma_{cap}(E_{\rm c.m.})}{\sigma^{noBU}_{cap}(E_{\rm c.m.})},
\label{P_BU}
\end{eqnarray}
we  extract the mean breakup  probability $<P_{BU}>(E_{\rm c.m.})$
averaged over all  partial waves $J$~(Fig.~7).
\begin{figure}
\includegraphics[scale=1]{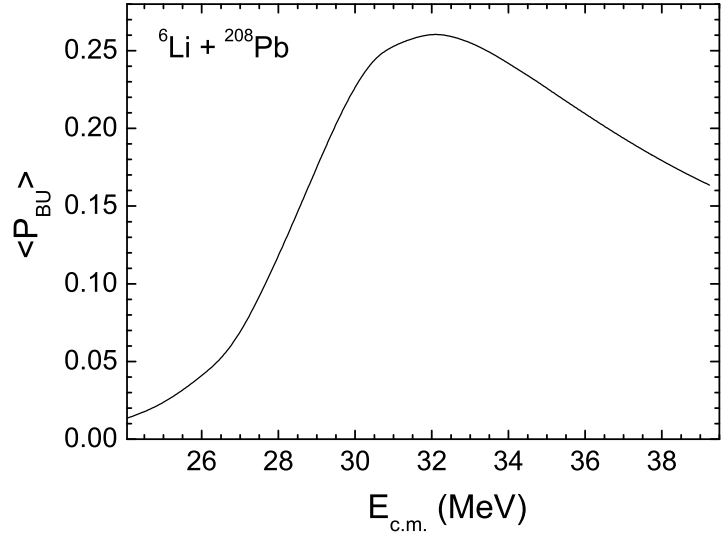}
\caption{
The extracted mean breakup probability $<P_{BU}>(E_{\rm c.m.})$ [Eq.~(14)]
as a function of bombarding energy $E_{\rm c.m.}$
for the $^{6}$Li+$^{208}$Pb reaction.
The used experimental quasielastic and quasielastic plus breakup data
are from Ref.~\protect\cite{Li6Pb}.
}
\label{7_fig}
\end{figure}
The value
of $<P_{BU}>$ has a maximum at $E_{\rm c.m.}-V_b\approx 4$ MeV ($<P_{BU}>$=0.26) and slightly (sharply)
decreases with increasing (decreasing) $E_{\rm c.m.}$. The experimental
breakup excitation function at backward angle
has the similar energy behavior~\cite{Li6Pb}.
By comparing  the calculated capture cross sections in the
absence of breakup and experimental capture (complete fusion)
data, the opposite energy trend is found in Ref.~\cite{Nash}, where
$<P_{BU}>$ has a minimum at $E_{\rm c.m.}-V_b\approx 2$ MeV  ($<P_{BU}>$=0.34) and globally increases
in  both sides from this minimum.
It is also shown in Refs.~\cite{Nash,PRSGomes4} that there are no systematic trends of breakup
in the complete fusion
reactions with the light projectiles $^{9}$Be, $^{6,7,9}$Li, and $^{6,8}$He at
near-barrier energies.
Thus, by employing the experimental quasielastic
backscattering, one
can obtain the additional information about the breakup process.
\begin{figure}
\includegraphics[scale=1]{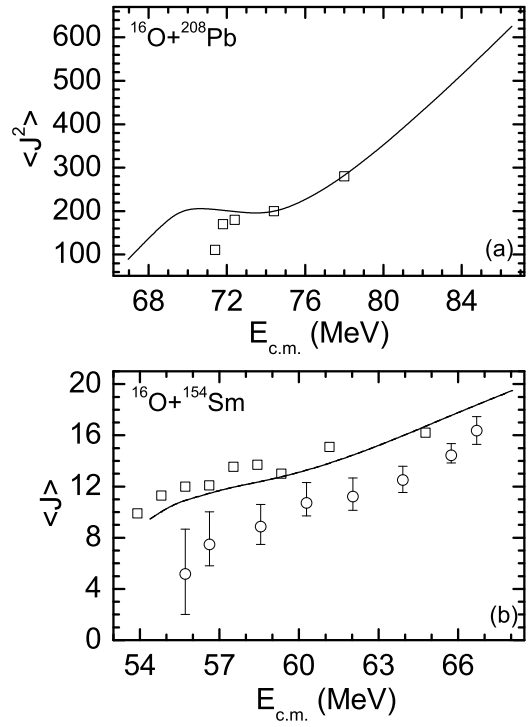}
\caption{
The extracted $<J>$ and $<J^2>$ for the  reactions $^{16}$O + $^{208}$Pb (a) and  $^{16}$O + $^{154}$Sm (b)
by employing Eqs.~(11) and  (12).
The used experimental quasielastic data are from Ref.~\protect\cite{Timmers2}.
The experimental  data of  $<J^2>$ and $<J>$ are from Refs.~\protect\cite{Vand} (open squares)
and~\protect\cite{Gil,Vand2} (open squares and circles), respectively.
}
\label{8_fig}
\end{figure}
\begin{figure}
\includegraphics[scale=1]{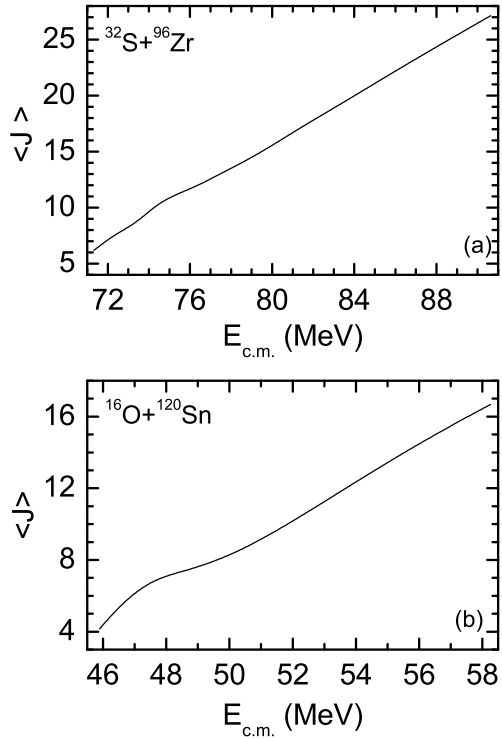}
\caption{
The extracted $<J>$  for the  reactions   $^{32}$S + $^{96}$Zr (a) and  $^{16}$O + $^{120}$Sn (b)
by employing Eq.~(11).
The used experimental quasielastic data are from Refs.~\protect\cite{Zhang3,Sinha}.
}
\label{9_fig}
\end{figure}
By using the Eqs.~(11) and (12) and experimental $P_{qe}$, we extract $<J>$ and $<J^2>$ of the captured system
for the  reactions $^{16}$O + $^{154}$Sm and $^{16}$O + $^{208}$Pb, respectively (Fig.~8).
The agreements with the results of direct measurements of the $\gamma-$multiplicities in
the corresponding complete fusion reactions are quite good.
For the $^{16}$O + $^{208}$Pb reaction at sub-barrier energies,
the difference between
the extracted and experimental angular momenta is related with the deviation of the
extracted capture excitation function  from the experimental one  (see Fig.~2).
In Fig.~9 we present the predictions
of $<J>$  for the  reactions $^{16}$O + $^{120}$Sn and $^{32}$S + $^{96}$Zr.

\section{Summary}

We realized that the found relationship between the quasielastic excitation function
and capture cross sections is working well, and
the quasielastic technique could be an  important and simple tool
in the study of the capture (fusion) research, especially, in the cold and hot   fusion
reactions and in the breakup reactions at energies near and above the Coulomb barrier.
Employing the quasielastic data, one can also extract the moments of the angular momentum
of the captured system.

We thank S.~Heinz, S. Hofmann and H.Q.~~Zhang for fruitful discussions and  suggestions. We are grateful to H.~Ikezoe,
C.J.~Lin, E.~Piasecki, and H.Q.~~Zhang for providing us their experimental data.
This work was supported by DFG, NSFC, RFBR, and  JINR grants.
The IN2P3(France)-JINR(Dubna) and Polish - JINR(Dubna)
Cooperation Programmes are gratefully acknowledged.
P.R.S.G. acknowledges the partial financial
support from CNPq and FAPERJ.\\


\end{document}